\documentclass[usenatbib]{mn2e}
\onecolumn

\usepackage{graphicx}
\usepackage{abbreviations}
\usepackage{amssymb}
\usepackage{amsmath}
\usepackage{color}

\newcommand{\bb}{\bmath{b}}
\newcommand{\bU}{\bmath{U}}

\newcommand{\bu}{\bmath{u}}

\newcommand{\bB}{\bmath{B}}

\newcommand{\etext}[1]{\quad\mbox{#1}\quad}

\newcommand{\Pd}[1]{\partial_{#1}}
\newcommand{\fracp}[2]{\left(\frac{#1}{#2}\right)}

\newcommand{\ort}[1]{ \bmath{i}_{#1} }

\newcommand{\sub}[1]{_{\mbox{\tiny #1}}}
\newcommand{\beq}{\begin{equation}}
\newcommand{\eeq}{\end{equation}}

\newcommand{\Km}{\mathcal K\sub{m}}
\newcommand{\Kf}{\mathcal K\sub{f}}
\newcommand{\Kq}{\mathcal K\sub{q}}
\newcommand{\sech}{\mbox{sech}}

\title[Tearing mode in two-fluid RMHD]{Relativistic Tearing and Drift-kink Instabilities in  
Two-fluid Simulations}
\author[Barkov \& Komissarov]
{
Maxim V. Barkov$^{1}$\thanks{E-mail: maxim.barkov@riken.jp (MVB)} and
Serguei S. Komissarov$^{2}$\thanks{E-mail: S.S.Komissarov@leeds.ac.uk (SSK)}\\ 
$^{1}$ Astrophysical Big Bang Laboratory, RIKEN, 351-0198 Saitama, Japan\\
$^{2}$ Department of Applied Mathematics, The University of Leeds, Leeds, LS2 9JT 
 }

\begin{document}
\date{Received/Accepted}
\maketitle

\begin{abstract} 

The stability of current sheets in collisionless relativistic pair plasma was studied
via two-dimensional two-fluid relativistic magnetohydrodynamic simulations with vanishing 
internal friction between fluids. In particular, we  investigated the linear growth of the 
tearing and drift-kink modes in the current sheets both with and without the guide field and   
obtained the growth rates which are very similar to what has been found in the corresponding PIC 
simulations. This suggests that the two-fluid simulations can be useful in studying the large-scale 
dynamics of astrophysical relativistic plasmas in problems involving magnetic reconnection.

\end{abstract}
                                                                                          
                                                                                          
\section{Introduction}
\label{introduction}

It is now well recognised that magnetic field is a ``major player'' in the dynamics of 
astrophysical plasma -- the Lorentz force shapes a wide variety of flows in the Universe. 
The dissipative effects are also important, leading to magnetic reconnection and explosive
release of stored magnetic energy. This could be of particular relevance in the astrophysics 
of neutron stars and black holes, which are expected to produce relativistic 
magnetically-dominated plasma. Magnetic reconnection accompanied by dissipation of 
magnetic energy may be the main processes leading to the observed non-thermal emission from
winds and jets produced by these compact relativistic objects 
\citep[e.g.][]{RL-92,DS-02,LB03,ZY-11,MU12,ssk-13,porth-13,porth-14}.      

The magnetic dissipation associated with the magnetic reconnection is not captured in the 
framework of ideal relativistic MHD, which is currently the most common tool of modelling  
astrophysical phenomena. The approximation of resistive MHD does introduce Ohmic dissipation 
of magnetic field but the astrophysical plasmas are often collisionless, whereas the 
resistivity has strong physical justification only for collisional plasmas.        

Kinetic models of plasma are better routed in fundamental physics and more 
suitable for collisionless plasma but they are also much more complex and computationally
expensive.  PIC-simulations, based on dynamics of individual particles  
(or rather ``super-particles''), are also quite expensive. Studies based on these methods shows 
that fast magnetic reconnection involves development of current sheets whose thickness 
is comparable to the electron skin depth, the kinetic scale absent in single fluid MHD 
\citep{ZH-01,ZH-07,BeBh12,CWU-14,SS-14,Liu-15}.

Half-way between these frameworks and the single fluid MHD are the multi-fluid models, 
where plasma is considered as a collection of several inter-penetrating charged and 
neutral fluids, coupled via macroscopic electromagnetic field.
Like a single fluid MHD, this approach is well suited for studying the large-scale 
dynamics of plasma flows. Moreover, it also captures some elements of plasma microphysics
in the form of collective interaction between its positively and negatively charged components,
which leads to the emergence of the plasma frequency and electron skin depth. 
Its generalised Ohm's law has several terms which introduce non-ideal 
properties even in the absence of explicit internal friction between fluids. 
For this reason, the multi-fluid approximation is considered as a potential alternative
to more expensive kinetic and PIC approaches when it comes to problems of macroscopic plasma 
dynamics where the magnetic reconnection plays an important dynamic role via restructuring 
of magnetic field and magnetic dissipation.    
For relativistic plasma, created in magnetospheres of neutron
stars and black holes via various pair production processes, a simple two-fluid
approximation involving electron and positron fluids may be sufficient.
Obviously, the lack of spectral information means that the fluid framework has rather 
limited potential for addressing such important issues as radiation and non-thermal particle 
acceleration.

So far, there has been only a rather limited effort to explore the potential 
of the two-fluid approximation in numerical modelling of relativistic plasma.   
\citet{ZHK09b,ZHK09a} used this approach for studying the
relativistic magnetic reconnection, \citet{Amano13} to study the termination
shocks of pulsar winds, and \citet{KO09} tried to construct two-fluid models of
steady-state pulsar magnetospheres. 
In the same way as this is done in resistive MHD simulations, \citet{ZHK09a,ZHK09b} 
used anomalous resistivity to trigger fast magnetic reconnection of Petschek-type. 
However, they noticed that the inertial terms of the generalised Ohm's law also make a 
significant contribution to the reconnection electric field, even exceeding that of 
the friction term, which represents the resistivity. Based on this observation, they 
suggested that the inertial terms alone may be sufficient to sustain magnetic 
reconnection. The robustness of this conclusion is not clear as 
they have also found that the simulations outcome strongly
depends on the model of resistivity. Moreover, they used the Lax-Wendroff numerical 
scheme which also introduces numerical resistivity, whose contribution to the 
reconnecting electric field exceeds the other terms \citet{ZHK09a,ZHK09b}. 
However, if the two-fluid model  can reproduce the reconnection rate sufficiently 
accurately then this approach becomes very useful for studying large-scale phenomena 
where magnetic restructuring and dissipation are important dynamical factors.         
 
Until recently, the fast magnetic reconnection was viewed in the context of the 
Petschek model with its compact diffusion zone, as opposed to the slow Sweet-Parker 
type reconnection of long and thin current sheets. 
However, long current sheets are unstable to tearing mode instability (TI), which splits it 
into much shorter current sheets separated by plasmoids. 
2D simulations discovered that at the non-linear stage the current sheet becomes highly dynamic, 
with mergers of original plasmoids and creation of new ones.   
This leads to a much higher overall reconnection rate \citep[e.g.][]{Biskamp1986,ST01,Lou07,BHY09,uzd-10}. 
In addition to TI, currents sheets are also subject to the so-called drift-kink 
instability (DKI) which grows faster \citep{ZH-07,CWU-14}. This discovery suggested that DKI 
may hinder the development of TI. However, recent 3D PIC simulations, where  both types of modes 
are allowed, show that TI is not suppressed and becomes dominant at the non-linear phase. 
The reconnection rates in 3D and 2D simulations are found to be similar 
\citep{SS-14, Liu-15}.        

Given the importance of TI and DK instabilities in the fast magnetic reconnection, 
the potential of the two-fluid model depends on how well it can describe their 
development. In this paper, we focus on the linear development of these instabilities
numerically. To this aim, we used our recently developed two-fluid code for pair plasma 
\citep[JANUS, ][]{barkov-14}. This code is based 
on a Godunov-type numerical scheme which is much less dissipative compared to the 
Lax-Wendroff one. It is third-order accurate in smooth regions, which makes it 
powerful tool for studying the instabilities. By setting the internal friction 
between the fluids (the resistivity) to zero we focus on the role of the inertial 
terms in the generalised Ohm's law. The results are compared with the growth
rates obtained via PIC simulations by other groups. In the follow-up paper, we will 
discuss the nonlinear phases of magnetic reconnection in the plasmoid-dominated regime.

\section{Two-fluid model of pair plasma}
\label{TFM}

Following \citet{ZHK09b} we adopt the 3+1 ($-+++$) Special Relativistic equations originated from
the covariant formulation by \citet{Gurovich86}.  The corresponding
dimensionless equations are \citep[for details see ][]{barkov-14}

\begin{itemize}

\item{the continuity equations}

\beq
\Pd{t}(n_\pm \gamma_\pm) + \nabla_i(n_\pm u_\pm^i)=0 \, ;
\label{continuity}
\eeq

\item{the total energy equation}

\beq
\Pd{t} \left( \sum\limits_{\pm} ( w_{\pm}\gamma_{\pm}^2 -p_{\pm}) +
            \frac{\Kq}{2\Km}(B^2+E^2) \right)
+ \nabla_i \left(  \sum\limits_{\pm} w_{\pm}\gamma_{\pm} u^i_{\pm} +
        \frac{\Kq}{\Km} e^{ijk}E_jB_k \right)= 0  \, ;
\label{total-energy}
\eeq

\item{the total momentum equation}

\beq
\Pd{t}\left( \sum\limits_{\pm} w_{\pm} \gamma_{\pm}  u_{\pm}^s +
   \frac{\Kq}{\Km} e^{sjk}E_jB_k   \right) +
\nabla_i \left( \sum\limits_{\pm}(w_{\pm} u^i_{\pm} u^s_{\pm} +p_{\pm}g^{is}) +
\frac{\Kq}{\Km}\left(-E^iE^s-B^iB^s + \frac{1}{2}(B^2+E^2)g^{is} \right)
\right)= 0\, ;
\label{total-momentum}
\eeq

\item{the Maxwell equations}

\begin{equation}
   \nabla_i B^i=0 \, ,
\label{Gauss_B}
\end{equation}

\begin{equation}
   \Pd{t} B^s + e^{sik}\Pd{i}E_k = 0 \, ,
\label{Faraday}
\end{equation}

\begin{equation}
\label{Gauss_E}
   \nabla_i E^i = \frac{1}{\Kq}(n_+\gamma_+ - n_-\gamma_-) \, ,
\end{equation}

\begin{equation}
   \Pd{t} E^s - e^{sik}\Pd{i}B_k = - \frac{1}{\Kq}(n_+u_+^s - n_-u^s_-) \, .
\label{Ampere}
\end{equation} 

\item{and the generalised Ohm's law}

\beq
\Pd{t}\left( \sum\limits_{\pm} \pm w_{\pm} \gamma_{\pm}  u_{\pm}^s \right) +
\nabla_i \left(
   \sum\limits_{\pm} \pm (w_{\pm} u^i_{\pm} u^s_{} +p_{\pm}g^{is} )
  \right)   =
      \frac{1}{\Km}\tilde{n} (E^s+e^{sik}v_i B_k) +
      \frac{2}{\Kf} n_+n_- (u^s_- - u^s_+)  \, .
\label{Ohm_m}
\eeq

\end{itemize}
In these equations $E$ and $B$ are the electric and magnetic field, $n_\pm$, $p_\pm$, 
$w_\pm$, $\gamma_\pm$ and $u_\pm=\gamma_\pm v_\pm$ are the density, pressure, 
relativistic enthalpy, Lorentz factor and 4-velocity of electron and positron fluids 
respectively, $g^{ik}$ is the spatial metric tensor of Minkowski space time and, 
$e^{ijk}$ is the Levi-Civita tensor, indexes s, i and k correspond to three spatial direction. 

In the Ohm's law, $\tilde{n} = n_+\gamma_+ + n_-\gamma_-$ is the total number density
of charged particles as measured in the laboratory frame and
$v^i=(n_+\gamma_+v^i_+ + n_-\gamma_-v^i_-)/\tilde{n}$ is their average
velocity in this frame. The last term of the Ohm's law describes the internal 
friction between the fluids, which is related to resistivity.    

The three dimensionless parameters in these equations are
\beq
\Kq = \frac{B_0}{4\pi e L_0 n_0} \,, \quad
\Km = \frac{m\sub{e}c^2}{e B_0 L_0} \,, \quad
\Kf = \frac{m\sub{e}c}{\varkappa\sub{f} n_0 L_0} \,,
\label{Ks}
\eeq
where $L_0$ is the characteristic length scale, the speed of light, $c$, is the 
characteristic speed, $B_0$ the characteristic value of magnetic (and electric)
field, $n_0$ the characteristic number density of particles and $e$ is the electron 
charge. The corresponding scales for the time is $T_0=L_0/c$, 
for the mass density  $m\sub{e} n_0$ and for the pressure and enthalpy $m\sub{e} c^2 n_0$,
$\varkappa\sub{f}$ is the dynamic coefficient
of friction between these fluids.

The dimensionless polytropic EOS is
\beq
w_\pm=n_\pm +\Gamma p_\pm /(\Gamma-1) \,.
\eeq
where $\Gamma$ is the ratio of specific heats.

In this paper we solve these equations numerically, using the code JANUS \citep{barkov-14}. 
The code is based on a conservative finite-difference scheme which utilises a third order
WENO interpolation \citep{LOC94,YaCa09} and a third order TVD time integration of the 
Runge-Kutta type \citep{ShOs88}, thus ensuring overall third order accuracy on smooth 
solutions. Hyperbolic fluxes are computed using the Lax-Friedrich prescription.  
The magnetic field is kept near divergence-free by means of the method of generalised 
Lagrange multiplier \citep{Munz00,Dedner02,ssk-rrmhd}.

\section{Harris current sheet} 
\label{HCS}

In the paper we study stability of the Harris current sheet using Cartesian coordinates 
aligned with the sheet.  In these coordinates, the initial magnetic field 
$\bB=(B_x(y),0,B_z)$, where $B_z$ is a uniform guide field and 

\beq
    B_x= B_\infty \tanh\fracp{y}{\delta}\,, 
\label{MF}
\eeq
where $\delta$ is the half-thickness of the current sheet and $B_\infty$ is the 
magnetic field strength far away from the sheet. The force equilibrium of the 
current sheet implies the total gas pressure distribution

\beq
   p_t=p_\infty + \frac{B_\infty^2}{8\pi} \left(1-\tanh^2\fracp{y}{\delta}\right) \,.
\label{TP}
\eeq
The gas pressure in the centre of the current sheet $p_0=p_\infty + B^2_\infty/8\pi$. 
Introducing the pressure ratio $f_p=p_0/p_\infty$, we find that

\beq
   f_p = 1+\beta_m^{-1}, 
\eeq   
where $\beta_m = 8\pi p_\infty/ B^2_\infty$ is the traditional (non-relativistic) magnetisation 
parameter of plasma.  Following previous studies, we assume that the plasma temperature is 
uniform and hence the particle density distribution follows that of the gas pressure.  

The half-thickness $\delta$ determines the drift speed of fluids in the current sheet. 
From the Faraday equation we find four velocity as 

\beq
    u_-^z = \frac{cB_\infty}{8\pi \delta e n_-}\sech^2\fracp{y}{\delta}\,,
\eeq
where we used the charge neutrality condition $n_-=n_+$, and hence $u_+=-u_-$. 

Using $B_\infty$, $\delta$ and $n^\pm_\infty$ as the characteristic scales $B_0$, $L_0$ 
and $n_0$ of the problem, we find 

\beq
    \Kq = 2 u_0 f_p \etext{and} \Km=\frac{1}{4\theta (f_p-1)}\,,
\eeq 
where $u_0$ is the magnitude of $u_-^z$ at the centre of the current sheet and 
$\theta=k_bT/m_ec^2$ is the dimensionless temperature, here $k_b$ is Boltzmann constant.
Following \citet{CWU-14}, we use $f_p=10$ and $u_0=0.75$ but set $\theta=10$ 
instead of $10^8$. The latter should not have a strong effect as in both cases 
the thermal energy dominates in the plasma inertia and this is indeed what 
has been found in the previous theoretical and numerical studies \citep{ZK-79,ZH-07}. 
Given these values, we have $\Kq=15$ and $\Km=0.042$.  The corresponding relativistic 
magnetisation parameter 

$$
   \sigma_\infty = \frac{B^2_\infty}{4\pi w_\infty} \approx 4.4\,, 
$$
where $w_\infty = w_{-,\infty}+w_{+,\infty}$. 
Using the definitions of the plasma Larmor radius, $\rho_0 =\theta m_e c^2/eB_\infty$, and 
the skin depth, $d_e^2 = \theta m_e c^2/(4\pi n_\infty e^2)$, 
as in \citep{CWU-14}, we find $\delta=2.4 \rho_0$ and $\delta=1.26 d_e$, 
which is similar to what they have in the setup of their PIC simulations 
($\delta=2.7 \rho_0$ and  $\delta=1.61 d_e$)\footnote{The difference is probably 
because they used $v_0=\sqrt{0.6}$ and not $v_0=0.6$ as stated in their paper. We 
realised this issue a bit too late.}.

\section{Simulations} 
\label{tests}

All simulations presented in this paper are two-dimensional (2D).
We split them in four groups. In Sections~\ref{sec-ti} and  \ref{sec-ki},  
we present our studies of the tearing and drift-kink instabilities of the Harris current 
sheet described in Sec.\ref{HCS}, without the guide field. The main goal is to 
obtain dispersion curves and compare them against the results of 
PIC simulations.  In Sec.\ref{sec-gf} we investigate the role of the guide field, by 
studying the response of modes with highest growth rates. In all models, 
the ratio of specific heats, $\Gamma=4/3$ and Courant number $\mbox{C} = 0.5$. All 
physical parameters are dimensionalised using the characteristic scales $c$, $L_0=\delta$, 
$B_0=B_\infty$ and $n_0=n_\infty$.   

In order to focus on the role of inertial terms in Ohm's law, we effectively remove 
the resistive term by setting $\Kf=10^{15}$. 

\begin{table*}
\caption{TI models. The case without guide field. Here $\lambda$ and $\omega\sub{i}$ are initial perturbation wavelength and perturbation grow rate respectively.}
\centering
\label{tab:tea}
\begin{tabular}{l|rrrrr}
\hline
Name & Resolution & domain X & domain Y & $\lambda$ & $\omega\sub{i}$ \\ 
\hline
TW06 & 128x128 &   [-3,3]     & [-5,5]    & 6 & 0.0\\
TW07 & 128x128 &   [-3.5,3.5] & [-5,5]    & 7 & 0.0 \\
TW08 & 128x128 &   [-4,4]     & [-5,5]   & 8 & 0.011\\
TW09 & 128x128 &   [-4.5,4.5] & [-5,5]   & 9 & 0.041\\
TW10 & 128x128 &   [-5,5]     & [-5,5]   & 10 & 0.0603 \\
TW11 & 128x128 &   [-5.5,5.5] & [-5,5]   & 11 & 0.077 \\
TW12 & 160x128 &   [-6,6]     & [-5,5]   & 12 & 0.085\\
TW14 & 160x160 &   [-7,7]     & [-7,7]   & 14 & 0.095\\
TW20 & 256x256 & [-10,10]     & [-10,10] & 20 & 0.100 \\
TW30 & 384x256 & [-15,15]     & [-10,10] & 30 & 0.089\\
TW40 & 512x256 & [-20,20]     & [-10,10] & 40 & 0.079\\
TW80 & 1024x512 & [-40,40]    & [-20,20] & 80 & 0.047\\
TW160 & 2048x1024 & [-80,80]  & [-40,40] & 160  & 0.030\\
\hline
TW10h & 192x192 &   [-5,5]    & [-5,5] & 10 & 0.0657\\
TW10H & 256x256 &   [-5,5]    & [-5,5] & 10 & 0.0669\\
\hline
\end{tabular}
\end{table*}

\subsection{Tearing instability without guide field}
\label{sec-ti}

For the study of the tearing instability, we consider a two-dimensional problem with 
$\Pd{z}=0$.  The current sheet is pushed out of equilibrium by perturbing  
the magnetic field, $\bB \to \bB+\bb$, where the  
divergence-free perturbation  
\beq
  \bb=b_0 e^{-(y/l)^2} \left[-\frac{2y}{(k l^2)} \cos(k x)\ort{x}+\sin(k x) \ort{y}\right] \,,
  \label{b0-tea}
\eeq    
where $k=2\pi/\lambda$ is the wavenumber and $b_0 = 10^{-3}$ is the amplitude of 
the perturbation. In the x direction, the size of the computational domain 
is set to be exactly one wavelength of the perturbation and we employ the periodic 
boundary conditions at the x boundaries. In the y-direction, we have a comparable size 
of the computational domain and use the free-flow boundary conditions.  
The basic parameters of the simulations are given in the Table~\ref{tab:tea}.

\begin{figure}
\centering
\includegraphics[width=80mm]{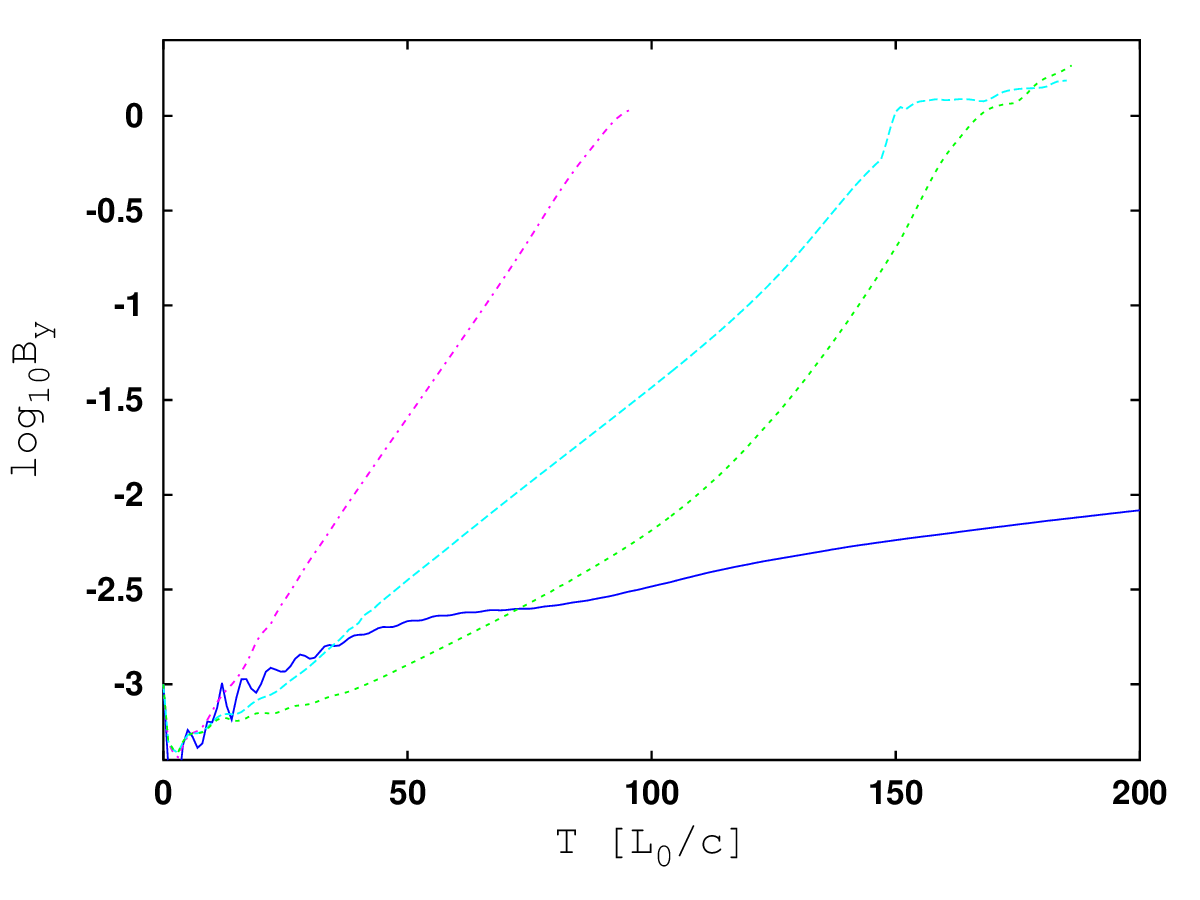}
\includegraphics[width=80mm]{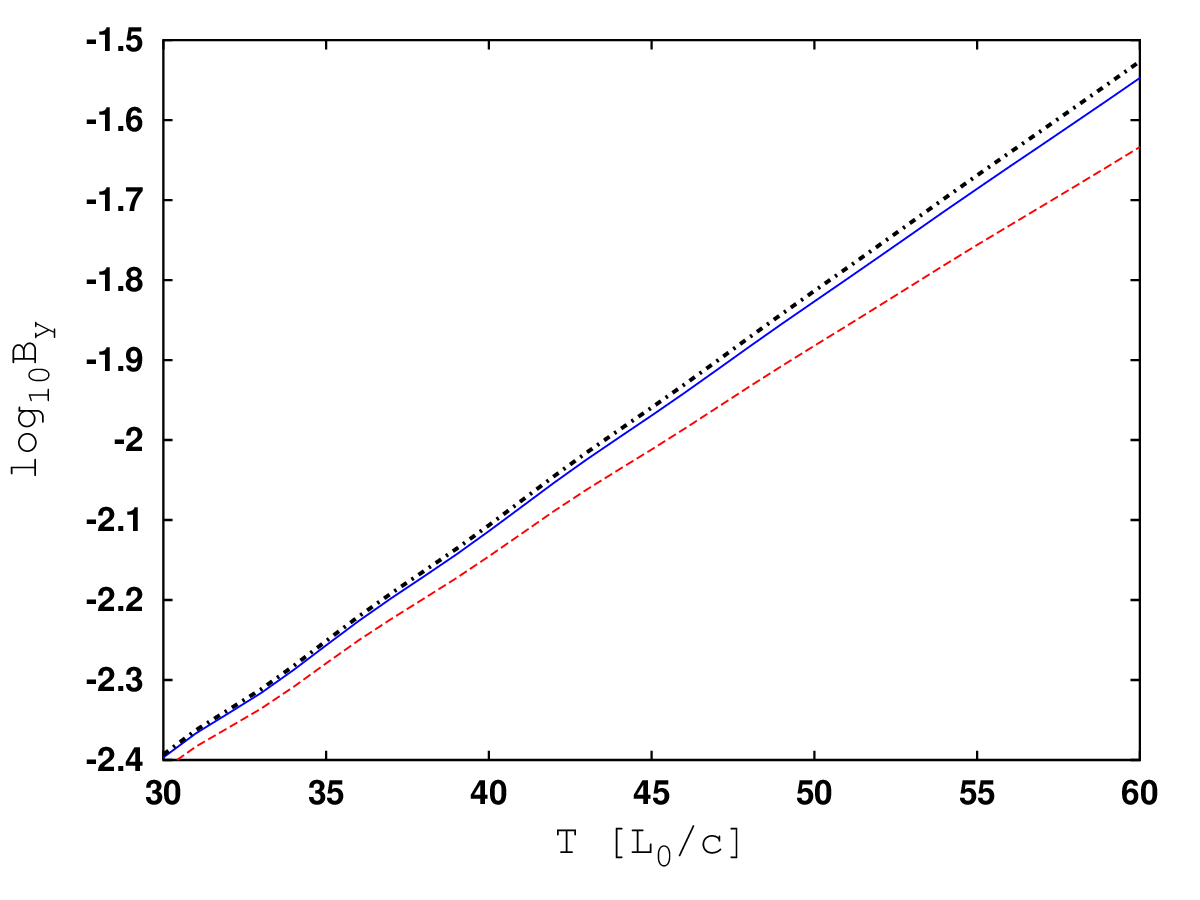}
\caption{
{\it Left panel:} Evolution of the perturbation amplitude for the models 
TW08 (blue solid line), TW30 (magenta dot-dashed line), 
TW80 (cyan dashed line) and TW160 (green doted line).
{\it Right panel:} Evolution of the perturbation amplitude for 
the models TW10 (solid line), TW10h (dashed line) and TW10H (dot-dashed line) which differ only by resolution.  
}
\label{fig:growth}
\end{figure}

\begin{figure*}
\centering
\includegraphics[width=7cm]{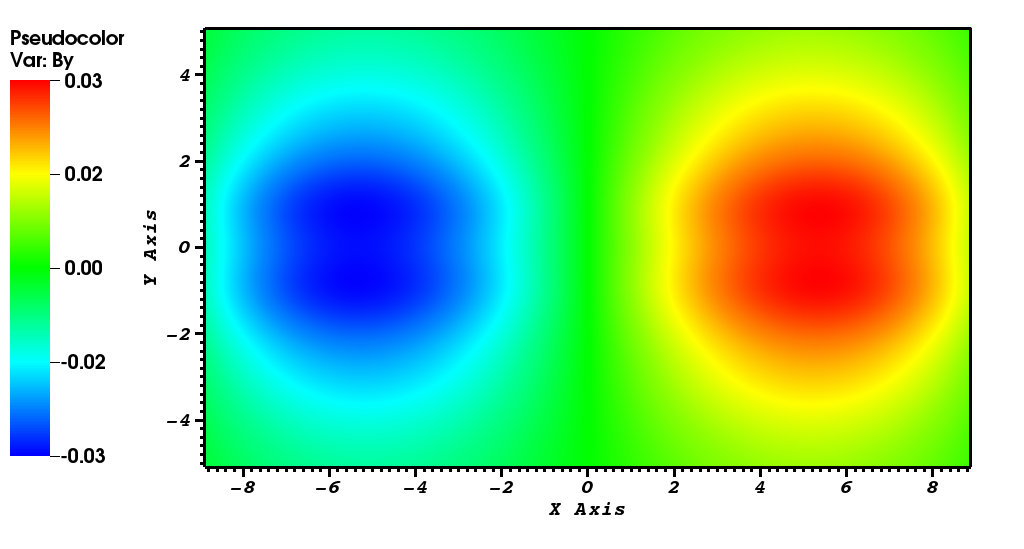}
\includegraphics[width=7cm]{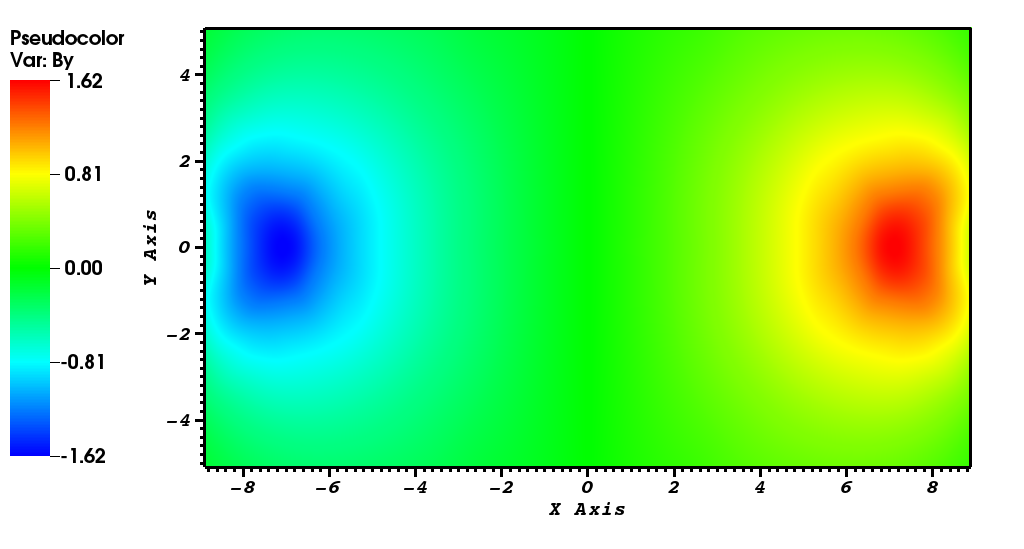}
\caption{
The distribution of $B_y$ for the model TW20 at the times  $t=23$ 
(top panel) and $t=88$ (bottom panel).  
}
\label{fig:short-w}
\end{figure*}

To quantify the perturbation amplitude we use the maximum value of $B^y$
in the computational domain. Figure~\ref{fig:growth} shows examples of the amplitude evolution for 
a number of models. As the initial perturbation is not a normal mode of the instability, it leads 
to excitement not only of the normal mode with the wavelength equal to the $x$ size of the computational 
box, the fundamental mode, but also its overtones as well as and propagating waves. The latter 
are partially transmitted through the $y$ boundaries of the computational box and do not grow in 
amplitude. Soon they become dominated by the unstable normal modes. When the wavelength of the fundamental 
mode is below the maximum of the dispersion curve, it completely dominates the evolution 
as the parasitic overtones grow slower. This is illustrated in Figure~\ref{fig:short-w}.   

\begin{figure}
\centering
\includegraphics[width=8cm]{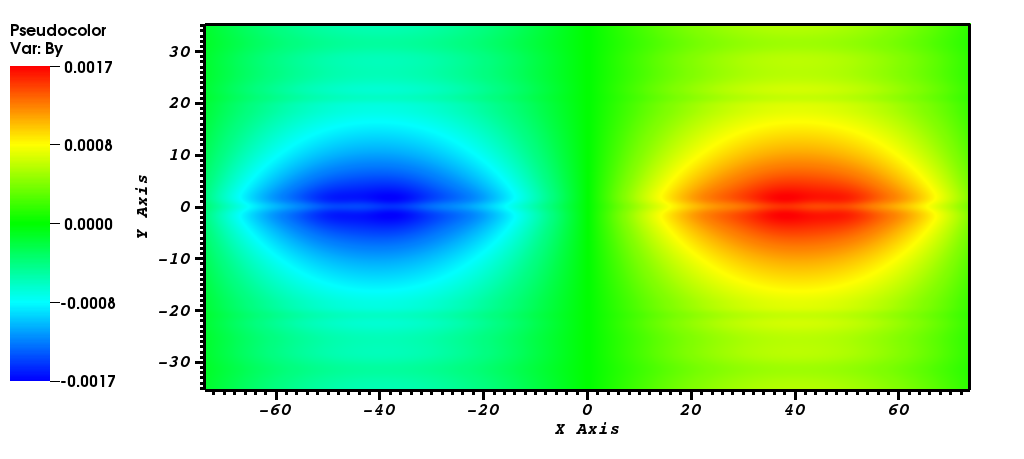}
\includegraphics[width=8cm]{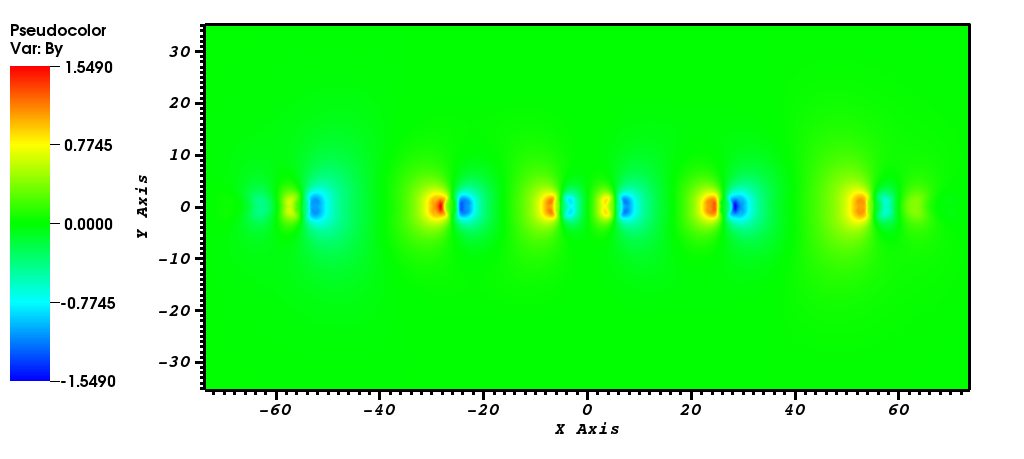}
\caption{
The distribution of $B_y$ for the model TW160 at the times  $t=60$ 
(top panel) and $t=180$ (bottom panel). One can see that initially it is the 
original perturbation of the wavelength equal to the domain length in the x direction
which dominates. However, at later times shorter wavelengths begin to dominate. 
}
\label{fig:long-w}
\end{figure}

When the wavelength of the fundamental mode is above the maximum, the evolution is more 
complicated. Initially, it dominates overtones simply because its initial amplitudes is higher. 
However, some parasitic overtones may now grow faster and eventually overtake it while still 
at the linear phase.  In the amplitude plots, this is manifested by an increase of 
the curve gradient, as exhibited by the curve of the TW160 model in Figure~\ref{fig:growth}. 
In the 2D plots of the solution, this is manifest by the appearance of dominant small 
scale structures (see Figure~\ref{fig:long-w}). This has to be taken into account when 
measuring the growth rate of the fundamental mode.

We have checked the convergence of our numerical results by comparing the data obtained 
with different numerical resolutions.  The right panel of Figure~\ref{fig:growth} shows 
the results for the model TW10 obtained with $128\!\times\!128$ cells, $192\!\times\!192$ cells 
(TW10h) and $256\!\times\!256$ (TW10H), which clearly indicate their convergence.   
Base on the study we conclude the numerical error of our growth rates does not exceed 10\%.

\begin{figure}
\centering
\includegraphics[width=80mm]{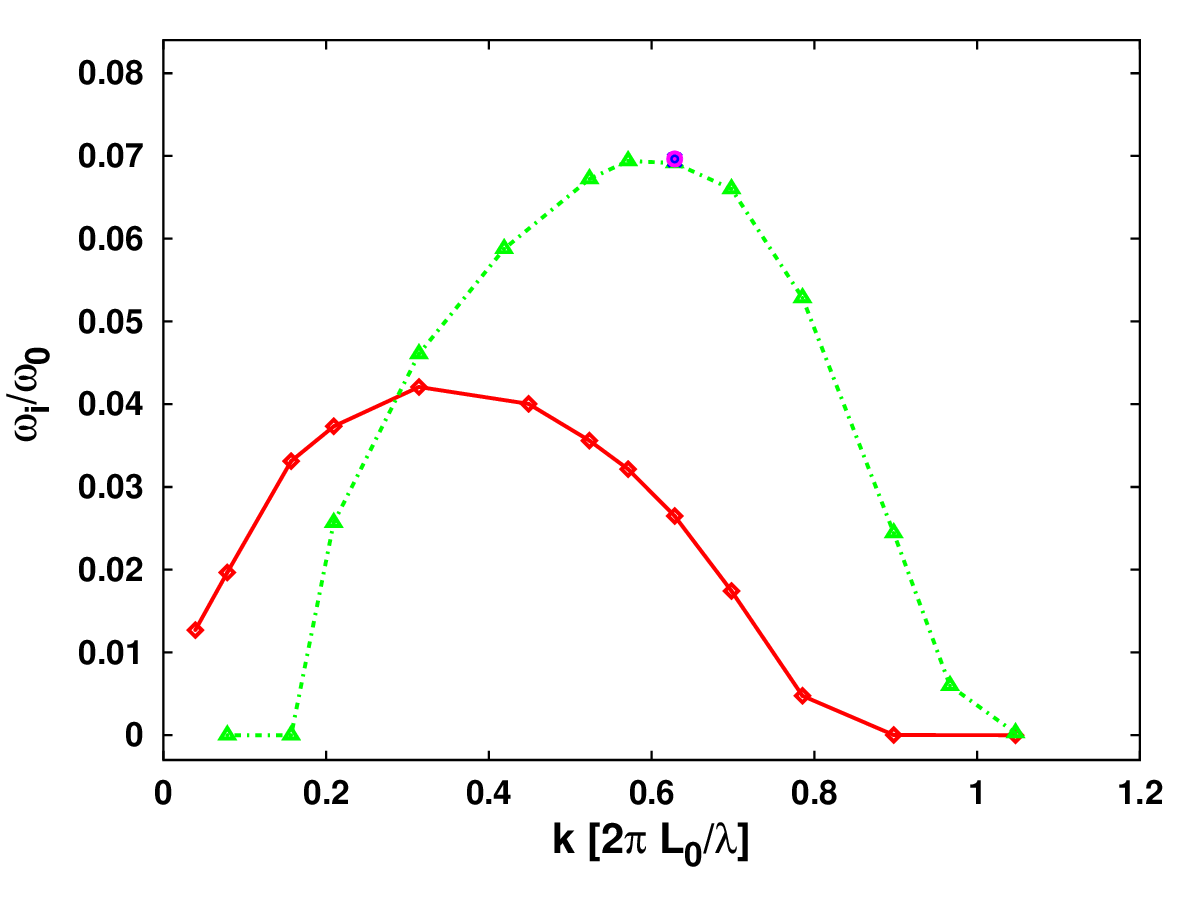}
\caption{
Growth rates of TI (thick red solid line) and DKI (green thick dot-dashed line) 
modes in the current sheet without guide field. The magenta circle shows the result 
obtained with doubled numerical resolution. 
}
\label{fig:ti-inc}
\end{figure}

The growth rates, $\omega_i$, of the fundamental modes are collected in Table~\ref{tab:tea} and 
Figure~\ref{fig:ti-inc}. The results agree with the theoretical models which 
predict instability for $0<k<1$ with a peak at $k\approx 0.5$ 
\citep{ZK-79,PK-07}. The  
PIC simulations show a broader dispersion curve, with unstable modes 
existing beyond $k=1$ and the peak growth rate  
$\omega_{i}/\omega_0\approx 0.045$, where $\omega_0 = \theta m_ec/eB$, 
at $k \approx 0.58$ \citep{CWU-14}. In our simulations, the peak is more 
pronounced and located at $k\approx0.3$. In order to compare our results, 
we note that with our scaling $\omega_0^{-1}=\theta\Km = 0.42$ and thus 
$\omega_{i}/\omega_0\approx 0.04$. Overall, we conclude that our results agree 
quite well with the PIC data.

\subsection{Drift-kink instability without guide field}
\label{sec-ki}

\begin{table*}
\caption{DKI models. The case without guide field. 
}
\begin{center}
\label{tab:kink}
\begin{tabular}{l|rrrrr}
\hline
Name & Resolution & Domain Z & Domain Y & $\lambda$ & $\omega\sub{i}$ \\
\hline
KW06 & 128x128 &   [-3,3] & [-5,5] & 6 & 0.0 \\
KW065 & 128x128 &  [-3.25,3.25] & [-5,5] & 6.5 & 0.0 \\
KW07 & 128x128 &   [-3.5,3.5] & [-5,5] & 7 & 0.058 \\
KW08 & 128x128 &   [-4,4] & [-5,5] & 8 & 0.135  \\
KW09 & 128x128 &   [-4.5,4.5] & [-5,5] & 9 & 0.159 \\
KW10 & 128x128 &   [-5,5] & [-5,5] & 10 & 0.165  \\
KW11 & 128x128 &   [-5.5,5.5] & [-5,5] & 11 & 0.1646 \\
KW12 & 192x128 &   [-6,6] & [-5,5] & 12 & 0.160 \\
KW15 & 192x128 &   [-7.5,7.5] & [-5,5] & 15 & 0.137 \\
KW20 & 256x128 & [-10,10] & [-5,5] &  20 & 0.110 \\
KW30 & 384x256 & [-15,15] & [-10,10] &  30 & 0.061 \\
KW40 & 512x256 & [-20,20] & [-10,10] &  40 & 0.0 \\
KW80 & 1024x512 & [-40,40] & [-20,20] &  80 & 0.0  \\
\hline
KW10h & 192x192 &   [-5,5] & [-5,5] & 10 & 0.16553  \\
KW10H & 256x256 &   [-5,5] & [-5,5] & 10 & 0.16581  \\
\hline
\end{tabular}
\end{center}
\end{table*}

For the study of the drift-kink instability, we consider a two-dimensional problem with
$\Pd{x}=0$.  The current sheet is pushed out of equilibrium by perturbing
the velocity field of both the electron and positron fluids, $\bU \to \bU+\bu$, where 
 
\beq
  \bu=u_0 \cos(k z)\ort{y} \,, 
  \label{b0-kink}
\eeq    
with $u_0=10^{-3}$. 
Like in the tearing simulations, the size of the computational domain in the z direction
is set to be exactly one wavelength of the perturbation and we employ relevant periodic
boundary conditions at the z boundaries. In the y-direction, we have a comparable size
and use the free-flow boundary conditions.  The basic parameters of the simulations
are given in the Table~\ref{tab:kink}.  

\begin{figure*}
\includegraphics[width=80mm]{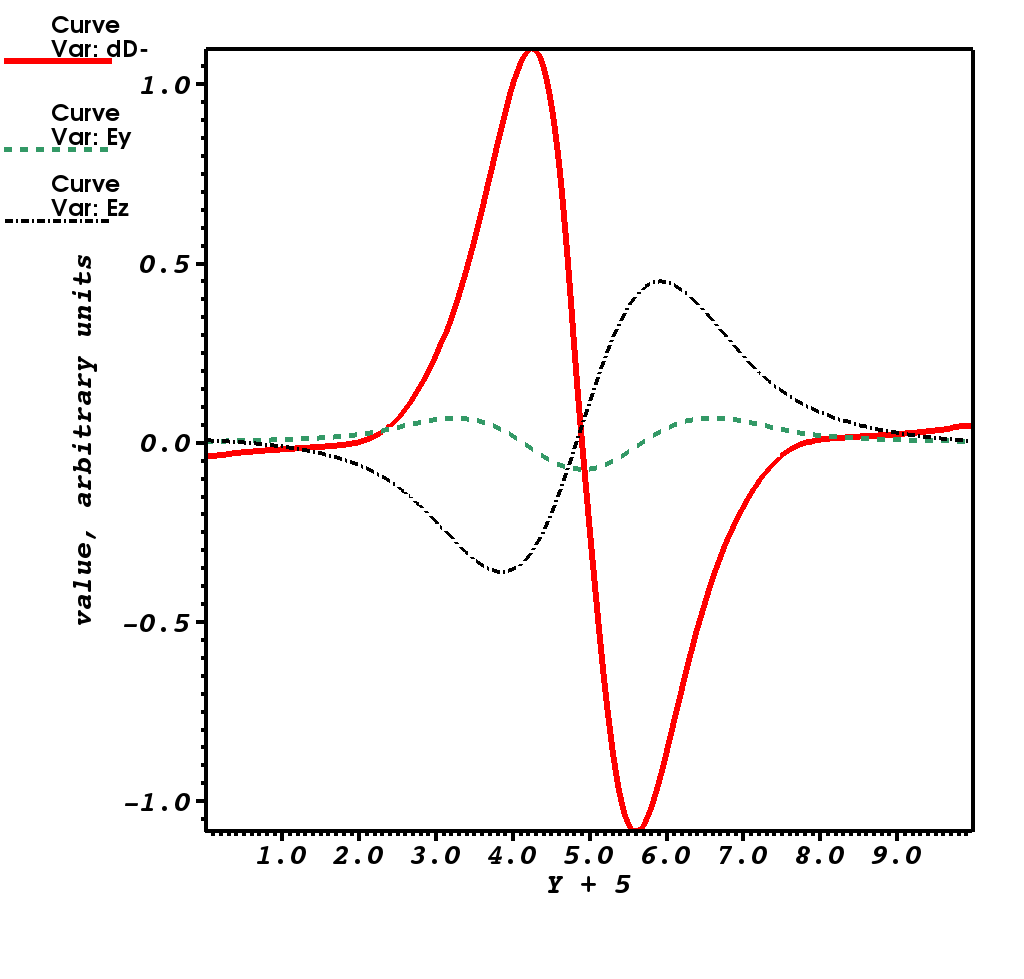}
\includegraphics[width=80mm]{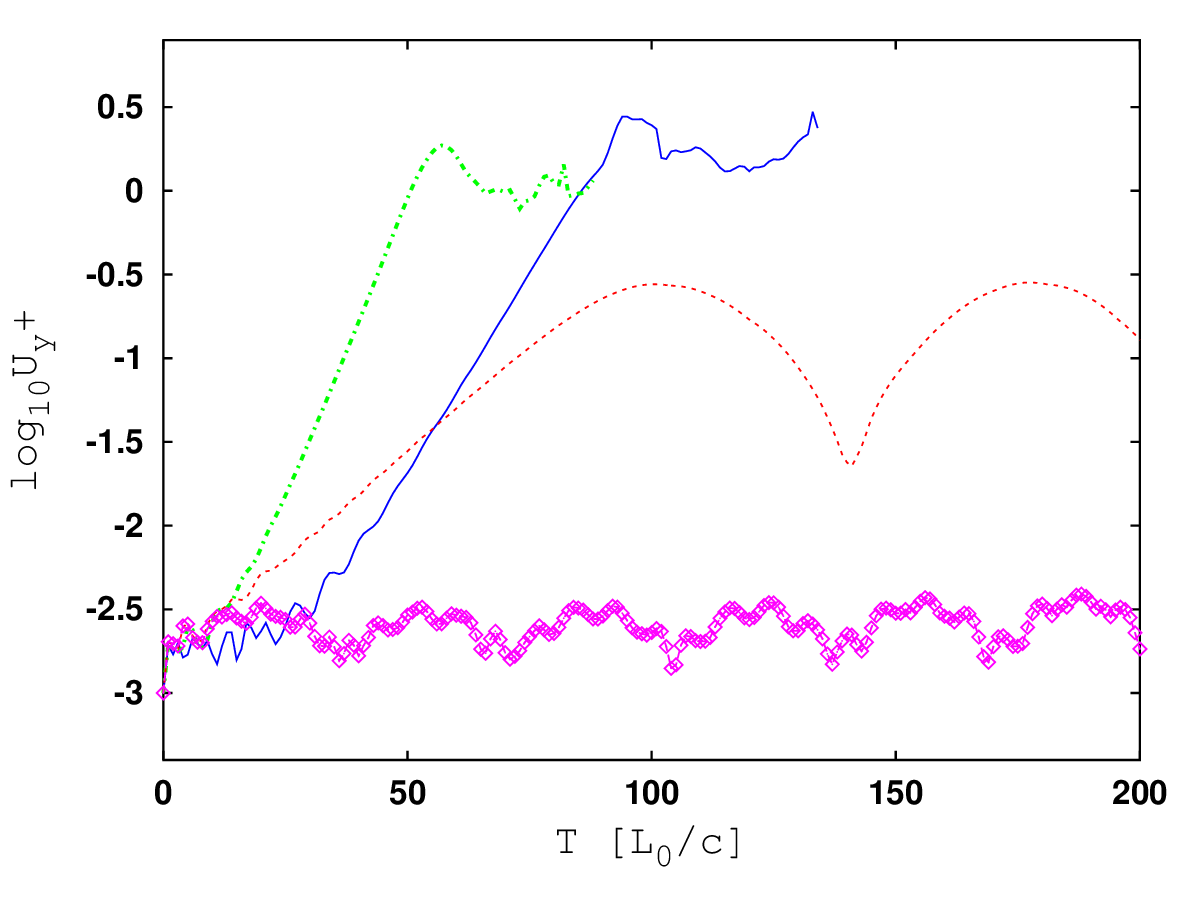}
\caption{{\it Left panel:} 
Structure of DK-modes without guide field. The lines show the perturbation of electron 
density (solid red line), $E_y\times 30$  (dashed green line) and $E_z\times 30$ 
(dash-dotted black line) as found in the model KW09.  The measurements are taken at $t=66$ 
along the line $z=2$.
{\it Right panel:}
Growth of DK modes without guide field. The curves represent models 
KW065 (magenta dashed line with diamonds), KW07 (red doted line), KW10 (green dot-dashed line) and 
KW20 (blue solid line). The shown quantity is the maximum value of $u^y_+$ in the computational box.   
}
\label{fig:tk-growth}
\end{figure*}

The left panel of Figure~\ref{fig:tk-growth} illustrates the structure of the unstable modes across 
the current sheet in our simulations. 
These results are in a good agreement with the structure of normal modes found in 
the linear theory of DK instability \cite{ZH-07}.  
We quantify the perturbation amplitude using the maximum value of $|u^y_+|$ in the computational 
box. The right panel of Figure~\ref{fig:tk-growth} shows typical examples of its evolution in 
the simulations. 

\begin{figure*}
\includegraphics[width=80mm]{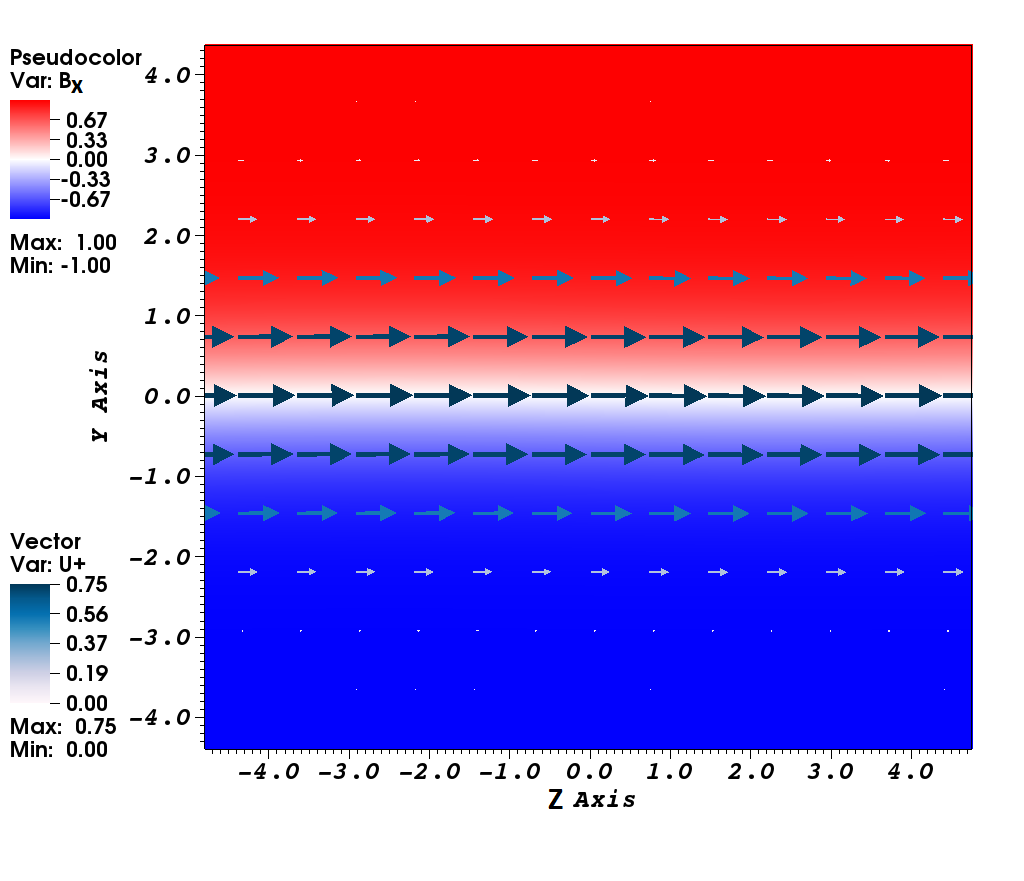}
\includegraphics[width=80mm]{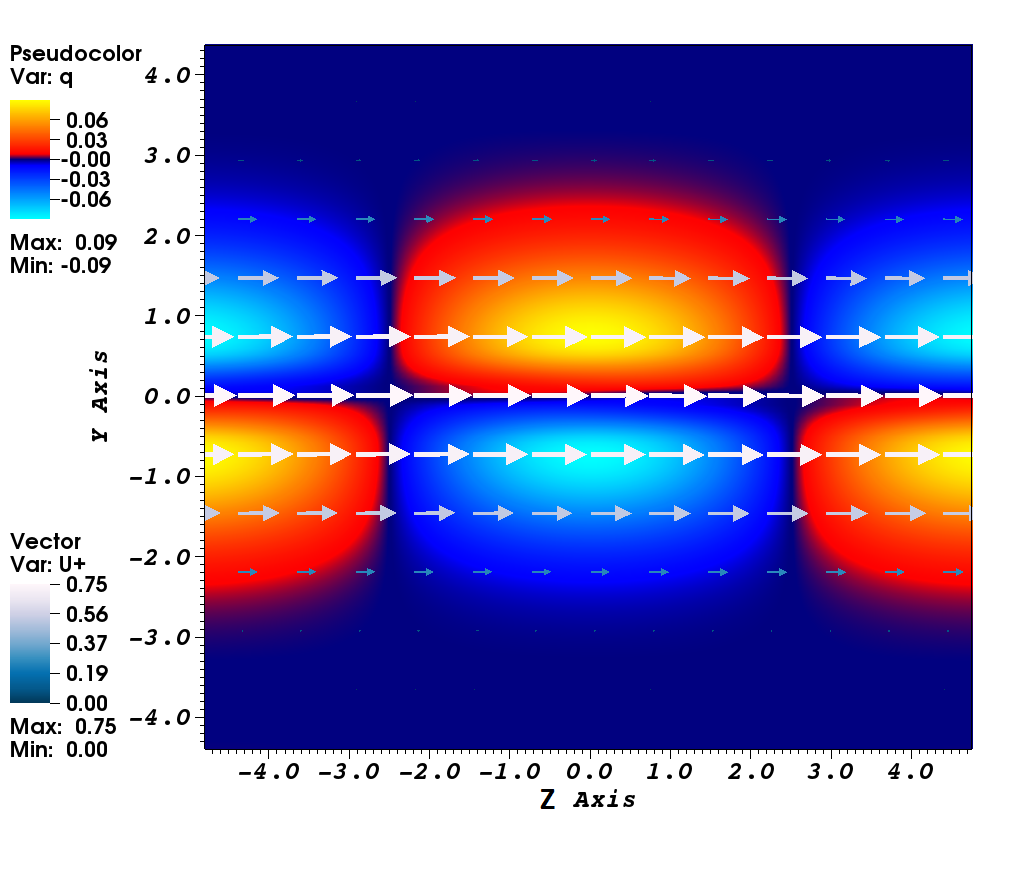}
\includegraphics[width=80mm]{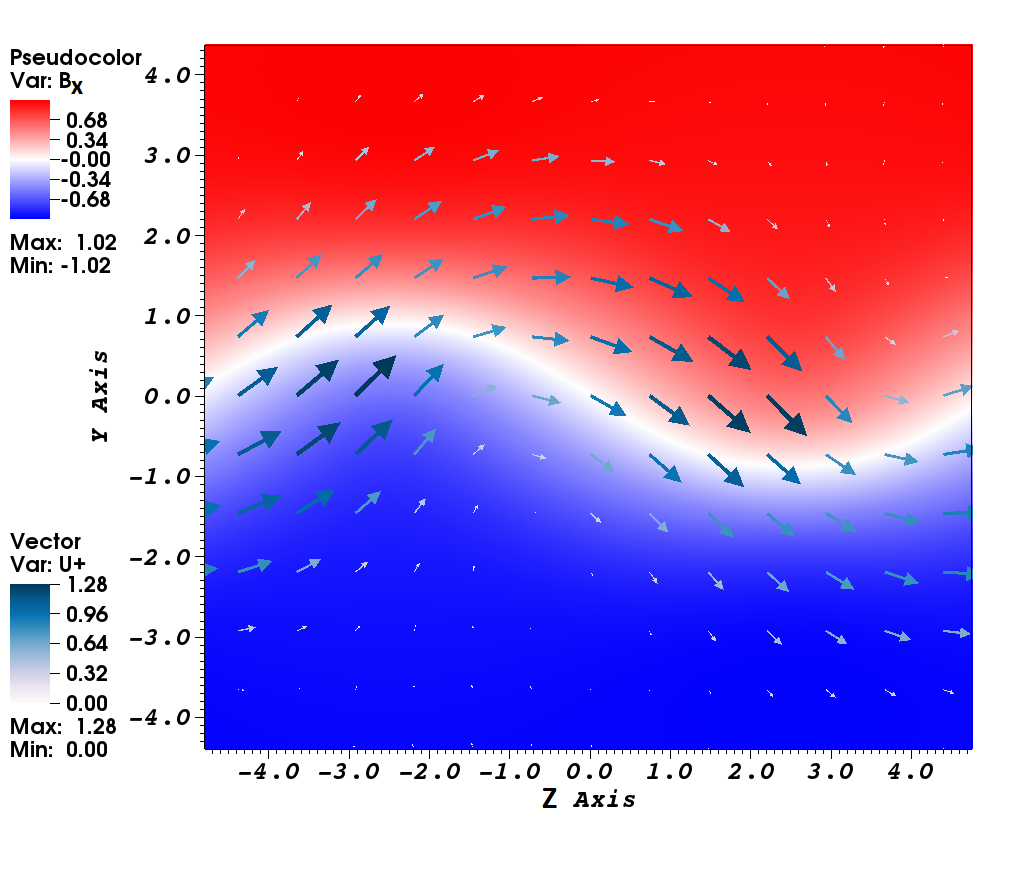}
\includegraphics[width=80mm]{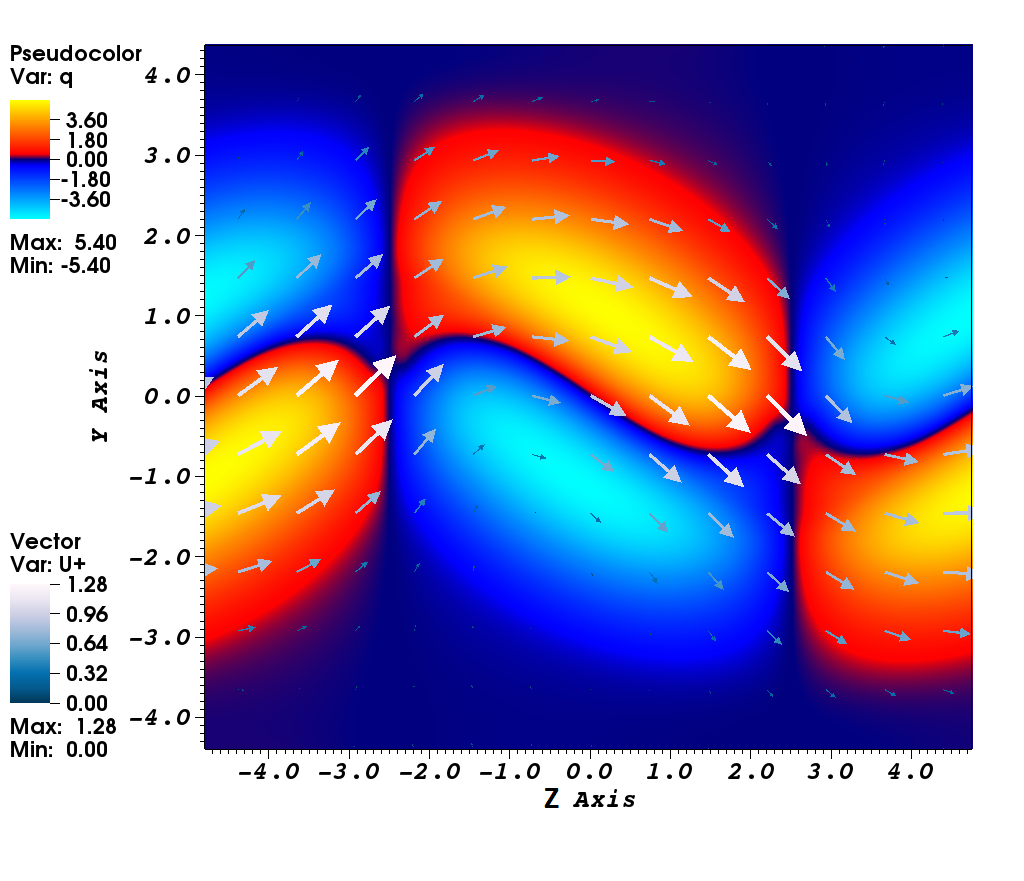}
\includegraphics[width=80mm]{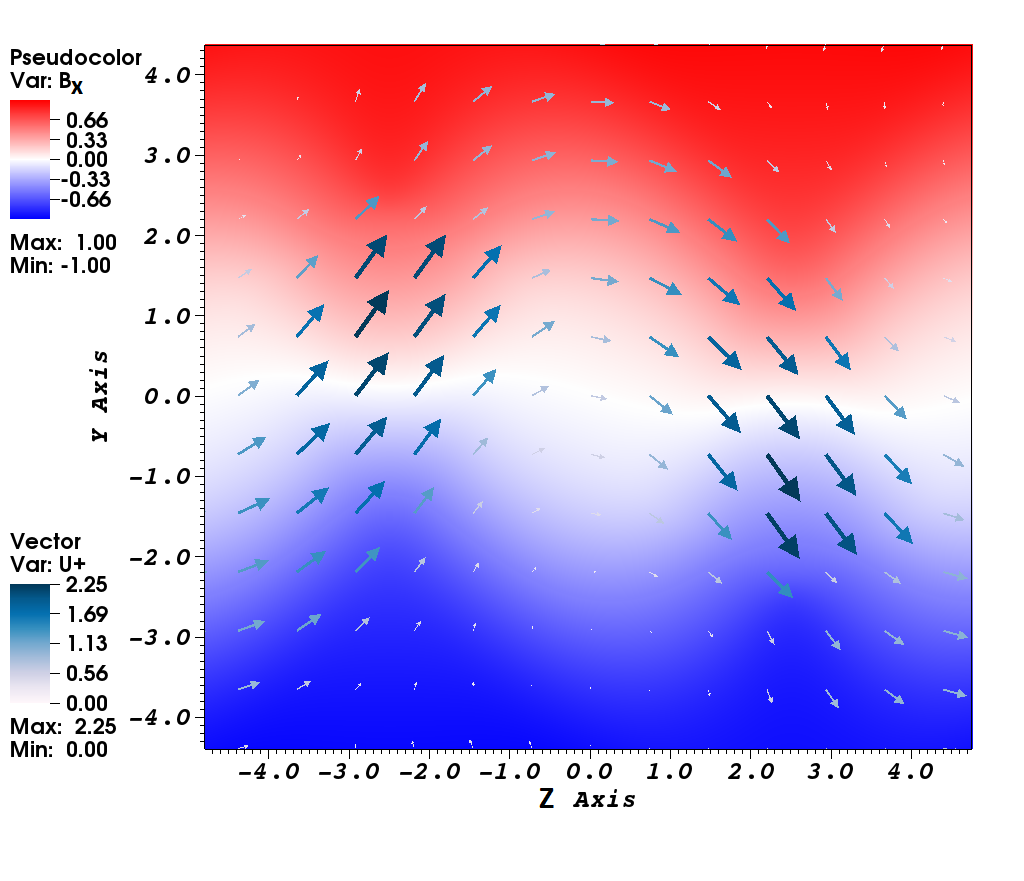}
\includegraphics[width=80mm]{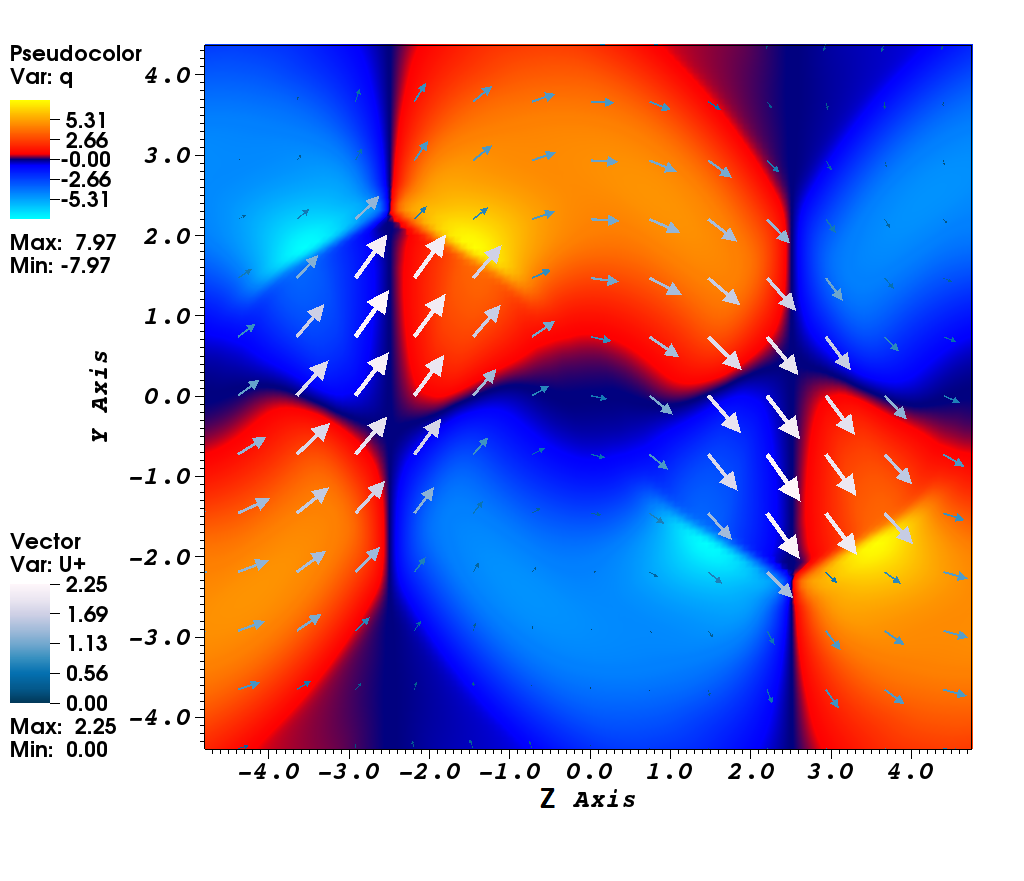}
\caption{Development of the drift-kink instability in the model KW10. In the left 
panels, the coloured image shows the distribution of the out-of-the-plane 
component of magnetic field $B_x$ and in the right panels, the distribution of 
electric charge. The arrows show the velocity field of positrons. 
The simulation time is $t=20,50$ and 58 (from top to bottom). By the time $t=58$, a 
significant fraction of magnetic energy has been dissipated and 
shock waves developed in the electron and positron fluids.    
}
\label{fig:tk-w1}
\end{figure*}

Figure~\ref{fig:ti-inc} shows the dispersion curve.  
Like in the tearing instability, the dispersion curve of DKI has 
a clear maximum and in the simulations with longer wavelengths, faster growing 
parasitic overtone modes can outperform the fundamental mode. 
In such cases, we compute $\omega_i$ only for the initial 
part of the amplitude curve, where the fundamental mode is still dominant. 
In order to verify that the numerical resolution is sufficient and the growth rates 
are trustworthy, we have carried out a separate convergence study. For example, we repeated 
the simulations KW10 with higher resolution: $192^2$ (model KW10h) and $256^2$ (model KW10H). 
The results indicate that the growth rate error for the model KW10 is below 1\% 
(see Table~\ref{tab:kink}). 

Like in the tearing instability, the unstable modes occupy the range $0<k<1$, though the 
long wavelength modes with $k<0.2$ appear to be suppressed (see Figure~\ref{fig:ti-inc}). 
In the PIC simulations the instability occurs even for $k >1$ but at a lower 
growth rate \citep{ZH-07,CWU-14}. In our simulations, the growth rate peaks at 
$k_{\rm max}\approx 0.6$, where it reaches the value $\omega_{\rm max}\approx 0.16$. 
Both \citet{ZH-07} whereas in the PIC simulations $k_{\rm max}\approx 0.7$ and  
$\omega_{\rm max}\approx 0.13$ \citep{ZH-07,CWU-14}. Thus the results of two-fluid and PIC 
simulations agree with each other quite well again.   
The linear analysis of \citet{ZH-07} shows that the instability domain extends
beyond $k=1$. However this theoretical results is not trustworthy as it is obtained using the 
long-wavelength approximation, $k\ll 1$. 

Interestingly, the short-wavelength modes appear to be non-decaying periodic or quasi-periodic
oscillations. The model KW065 is one such example. Its amplitude remains on the level 
of initial perturbation. The model KW07 seems to be a transitional case, where the initial 
phase of exponential growth terminates at a relatively low amplitude and is followed 
by oscillations. 

In the PIC simulations, the non-linear phase of DKI is characterised by magnetic dissipation, 
plasma heating, and widening of the current sheet. All these properties are observed in our 
simulations as well. Moreover, we find that shock waves develop in electron and positron fluids 
( see Figure~\ref{fig:tk-w1}) and they play an important role in plasma heating.

\subsection{Current sheets with guide field}
\label{sec-gf}

Following \citet{ZH-08} and \citet{CWU-14} we first study the effect of guide field on 
the fastest growing modes in the case without the the guide field.  
In our study, these are $k\approx0.31$ ($\lambda=20$) for TI and 
$k\approx 0.63$ ($\lambda=10$) for DKI.  
The computational domain is $[-5,5]\times[-5,5]$ with $128\times128$ cells for the 
TI simulations and $[-10,10]\times[-5,5]$ with $256\times128$ cells 
for the DKI simulations. The strength of the guide field is described by 
the parameter $\alpha\sub{gf} = B_z/B_\infty$. 
The perturbations are introduced in exactly the same way as in the models 
without the guide field. 

\begin{center}
\begin{table*}
\caption{The full set of models with guide field in the study of DKI.}
\label{tab:kinkgaf}
\begin{tabular}{l|rrrlrr}
\hline
Name & Resolution & Domain Z & Domain Y & $\lambda$ & $\alpha_{\rm gf}$  & $\omega_i$  \\
\hline
Ka09W05  & 192x192 &   [-2.5,2.5]   & [-5,5] & 5   & 0.9 & 0.0 \\
Ka09W055 & 192x192 &   [-2.75,2.75] & [-5,5] & 5.5 & 0.9 & 0.064 \\
Ka09W06  & 192x192 &   [-3.0,3.0]   & [-5,5] & 6   & 0.9 & 0.107 \\
Ka09W07  & 192x192 &   [-3.5,3.5]   & [-5,5] & 7   & 0.9 & 0.128 \\
Ka09W08  & 192x192 &   [-4.0,4.0]   & [-5,5] & 8   & 0.9 & 0.113 \\
Ka09W09  & 192x192 &   [-4.5,4.5]   & [-5,5] & 9   & 0.9 & 0.092 \\
Ka09W10  & 192x192 &   [-5.0,5.0]   & [-5,5] & 10  & 0.9 & 0.054 \\
Ka09W12  & 192x192 &   [-6.0,6.0]   & [-5,5] & 12  & 0.9 & 0.0 \\
\hline
Ka12W04  & 192x192 &   [-2.0,2.0]   & [-5,5] & 4   & 1.2 & 0.0 \\
Ka12W045 & 192x192 &   [-2.25,2.25] & [-5,5] & 4.5 & 1.2 & 0.0 \\
Ka12W05  & 192x192 &   [-2.5,2.5]   & [-5,5] & 5   & 1.2 & 0.079 \\
Ka12W055 & 192x192 &   [-2.75,2.75] & [-5,5] & 5.5 & 1.2 & 0.106 \\
Ka12W06  & 192x192 &   [-3.0,3.0]   & [-5,5] & 6   & 1.2 & 0.104 \\
Ka12W065 & 192x192 &   [-3.25,3.25] & [-5,5] & 6.5 & 1.2 & 0.086 \\
Ka12W07  & 192x192 &   [-3.5,3.5]   & [-5,5] & 7   & 1.2 & 0.0 \\
\hline
Ka16W038 & 192x192 &   [-1.9,1.9]   & [-5,5] & 3.8 & 1.6 & 0.0 \\
Ka16W04  & 192x192 &   [-2.0,2.0]   & [-5,5] & 4   & 1.6 & 0.037 \\
Ka16W045 & 192x192 &   [-2.25,2.25] & [-5,5] & 4.5 & 1.6 & 0.068 \\
Ka16W05  & 192x192 &   [-2.5,2.5]   & [-5,5] & 5   & 1.6 & 0.050 \\
Ka16W055 & 192x192 &   [-2.75,2.75] & [-5,5] & 5.5 & 1.6 & 0.017 \\
Ka16W06  & 192x192 &   [-3.0,3.0]   & [-5,5] & 6   & 1.6 & 0.0 \\
\hline
\end{tabular}
\end{table*}
\end{center}

\begin{figure}
\includegraphics[width=80mm]{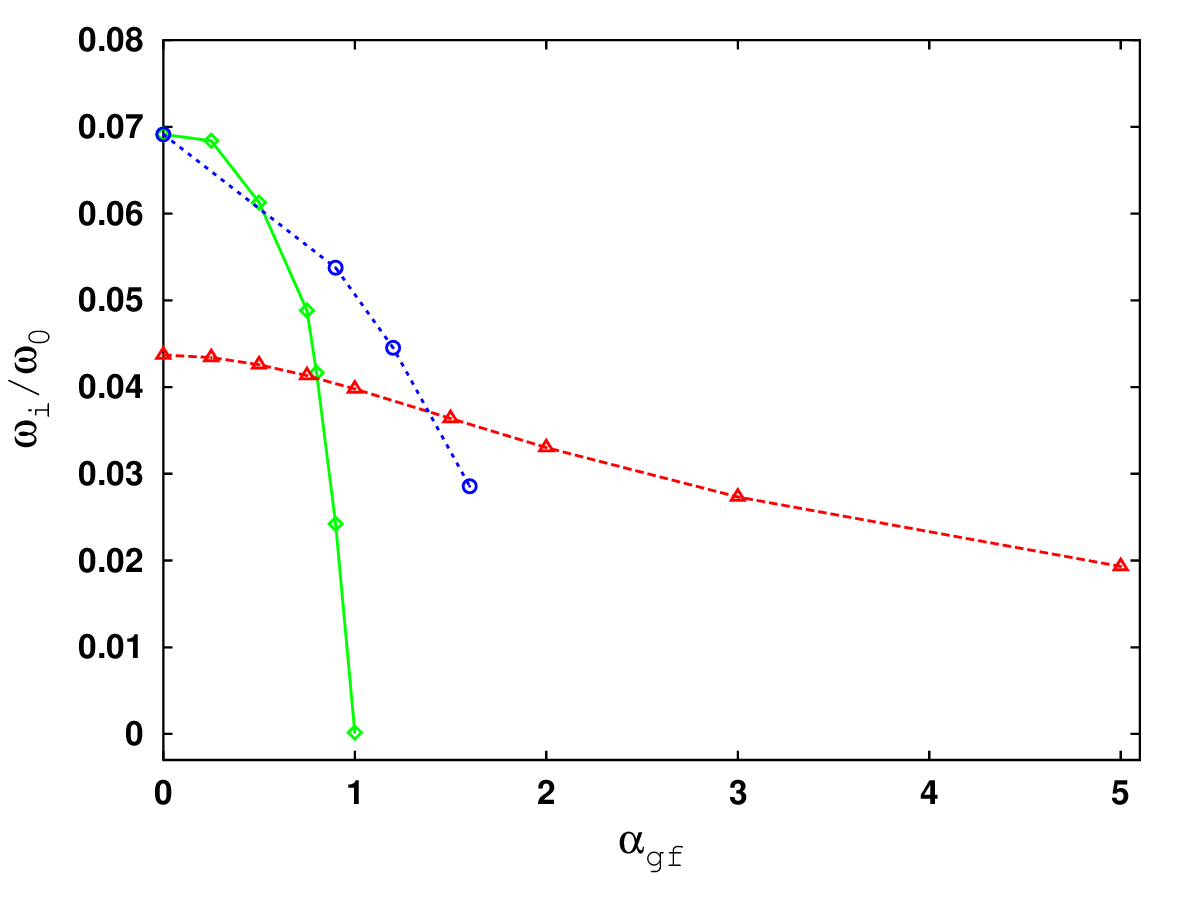}
\includegraphics[width=80mm]{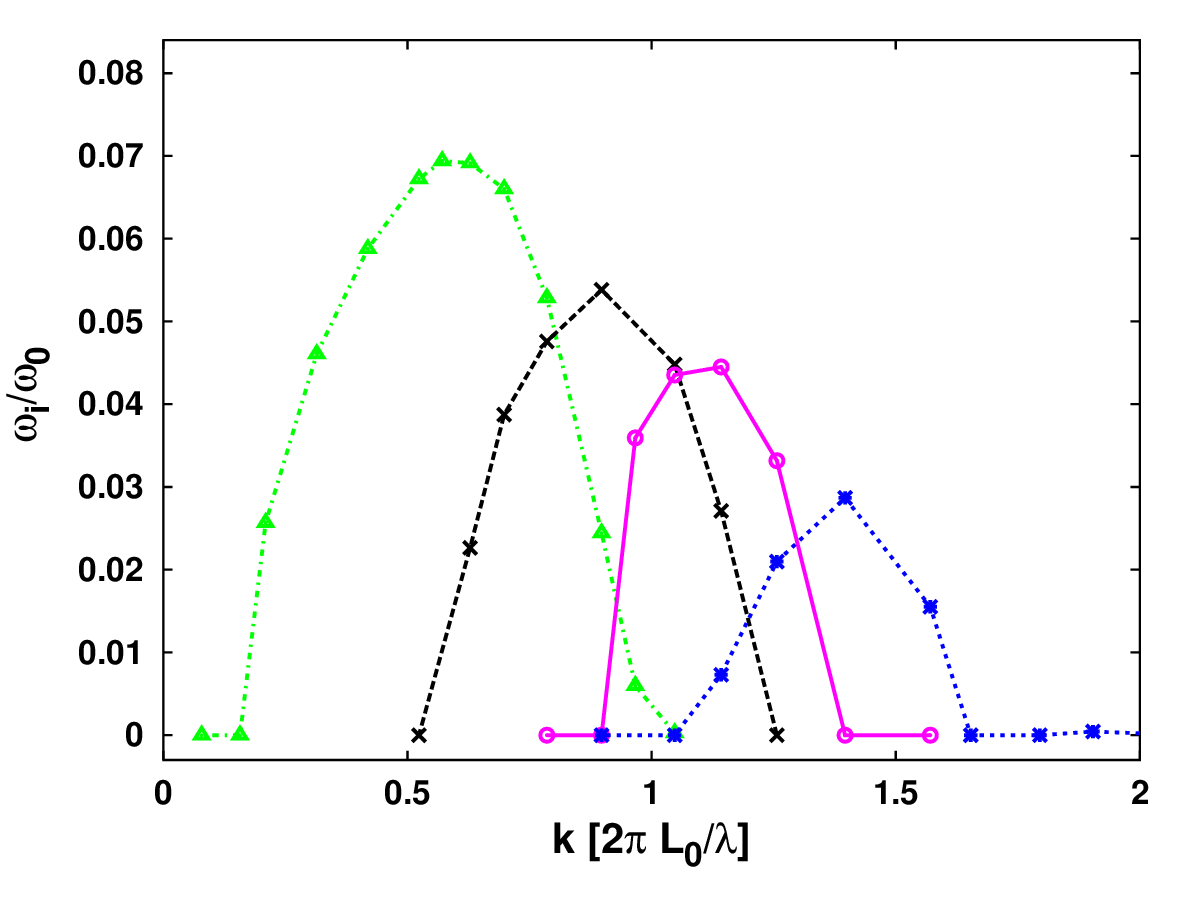}
\caption{
{\it Left panel}: Dependence of the growth rates of the tearing and drift-kink instabilities 
on the strength of guide field, $\alpha_{\rm gf} = B_z/B_\infty$. The red line shows 
the TI mode with $k\approx0.3$ ($\lambda=20$) and the green line the DKI mode 
with $k\approx 0.6$ ($\lambda=10$). The blue line shows the growth rate for the fastest 
growing DKI mode (its wavelength depends on $\alpha_{\rm gf}$).    
{\it Right panel:} 
Growth rates of the drift-kink instability in the presence of the guide field. 
The lines are the dispersion curves for $\alpha_{\rm gf} = 0$ (green triangles), 
0.9 (black crosses), 1.2 (magenta squares) and 1.6 (magenta stars).
}
\label{fig:gf}
\end{figure}

The results are shown in Figure~\ref{fig:gf}. As in the previous studies, the guide field makes a stronger 
impact on the DKI mode than on the TI mode. 
For the TI mode, we find that the growth rate is reduced by 50\% only at 
$\alpha\sub{gf}=5$, which is in agreement with the two-fluid linear analysis by \citet{ZH-08} 
and their PIC simulations.   
The PIC simulations by \citet{CWU-14} show a somewhat stronger effect, with a 45\% 
reduction already at $\alpha\sub{gf}=1$. However, their curve is not monotonic, which may 
indicate higher numerical  errors. As to the DKI mode, we find that it is totally suppressed 
when $\alpha\sub{gf}>1$.  This is in a good agreement with the linear stability analysis 
of \citet{ZH-08}, who find that for the DKI mode with $k=0.7$ the critical guide field is  
$\alpha\sub{gf,c}\approx0.5$, which also agrees with the results of their 
PIC simulations. Based on their PIC simulations, \citet{CWU-14} find  $\alpha\sub{gf,c}\approx0.8$ 
for $k=0.67$, which is even closer to our results. 

Given the strong effect of the guide field on the drift-kink instability, we have carried out 
additional simulations with the aim to clarify the dependence of the DKI dispersion curve on 
the guide field strength. The parameters of these simulations are given in 
Table~\ref{tab:kinkgaf} and their results are illustrated in Figure~\ref{fig:gf}.
The surprising result is that the growth rate is not uniformly reduced for all wavelengths.  
As the guide field increases, the peak of the curve does get lower but in addition 
the unstable range shifts towards shorter wave lengths. As a result, some modes which grow 
at $\alpha\sub{gf}=0$ become completely stabilised for $\alpha\sub{gf}\not=0$ and the 
other way around (see the right panel of Figure~\ref{fig:gf}). The blue line in the left panel of 
Figure~\ref{fig:gf} shows the dependence of the maximal growth rate on $\alpha\sub{gf}$.

\section{Conclusion}

In this work we studied the tearing and drift-kink instabilities of 
current sheets in collisionless electron-positron plasma by means of 
2D two-fluid computer simulations. We set the internal friction (resistivity) 
to zero and considered current sheets of thickness comparable to the 
electron skin depth, so that the inertial terms of the generalised Ohm's law
become significant. Our results are compared with those of the PIC simulations 
carried out by other researches for current sheets with similar parameters. 
We find that there is a good overall agreement between the two-fluid and 
PIC simulations. In both cases, the fastest growing modes have 
very similar wavelengths and growth rates when the guide field is small. 
In both cases, the guide field reduces the growth rates of unstable modes. 
There are some differences too. For example, the unstable range of both TI and DKI 
appears to be somewhat narrower and the guide field has a weaker stabilising effect 
on the TI mode in the two-fluid simulations. 
We also find that, in addition to getting lower, the dispersion curve of DKI 
also shifts towards higher wavenumbers when the guide fields gets stronger. 
We cannot say if this is in agreement with the PIC simulations due to the lack 
of relevant PIC data. 

It would be naive to hope that the two-fluid simulations could exactly reproduce the 
results of PIC simulations, and they do not. However, the differences appear to be rather 
minor. This suggests that the two-fluid model can adequately describe the macroscopic 
dynamics of plasma with collisionless currents sheets, yielding sufficiently accurate 
magnetic reconnection rates.  
In order to confirm this we have started a study of 2D magnetic reconnection in 
the plasmoid dominated regime. The preliminary results are encouraging.

\section{Acknowledgments}
The simulations have been carried out either on 
a work station  with multi-core processors or/and on the CFCA cluster (XC30) of 
National Astronomical Observatory of Japan. 
We make 2D visualisation and analysis using package VisIt \citep{HPV:VisIt} and Octave.
SSK acknowledges support by STFC under the standard grant EP/N023986/1.
BMV acknowledge partial  support  by  JSPS (Japan Society for the Promotion of Science):
No.2503786, 25610056, 26287056, 26800159. 
BMV also  MEXT (Ministry of Education, Culture, Sports, Science and Technology):
No.26105521 and RFBR  grant  12-02-01336-a.


\bibliographystyle{mn2e}

\end{document}